\documentclass[11pt]{article}
\usepackage{graphicx}

\begin{document}

\def\xslash#1{{\rlap{$#1$}/}}
\def \p {\partial}
\def \dd {\psi_{u\bar dg}}
\def \ddp {\psi_{u\bar dgg}}
\def \pq {\psi_{u\bar d\bar uu}}
\def \jpsi {J/\psi}
\def \psip {\psi^\prime}
\def \to {\rightarrow}
\def\bfsig{\mbox{\boldmath$\sigma$}}
\def\DT{\mbox{\boldmath$\Delta_T $}}
\def\xit{\mbox{\boldmath$\xi_\perp $}}
\def \jpsi {J/\psi}
\def\bfej{\mbox{\boldmath$\varepsilon$}}
\def \t {\tilde}
\def\epn {\varepsilon}
\def \up {\uparrow}
\def \dn {\downarrow}
\def \da {\dagger}
\def \pn3 {\phi_{u\bar d g}}

\def \p4n {\phi_{u\bar d gg}}

\def \bx {\bar x}
\def \by {\bar y}

\begin{center}
{\Large\bf Soft-Gluon-Pole Contribution in Single Transverse-Spin Asymmetries of Drell-Yan Processes }
\par\vskip20pt
J.P. Ma$^{1,2}$  and H.Z. Sang$^{3}$    \\
{\small {\it
$^1$ Institute of Theoretical Physics, Academia Sinica,
P.O. Box 2735,
Beijing 100190, China\\
$^2$ Center for High-Energy Physics, Peking University, Beijing 100080, China  \\
$^3$ Institute of Modern Physics,
School of Science,
East China University of Science and Technology,
130 Meilong Road, Shanghai 200237, P.R. China
}} \\
\end{center}
\vskip 1cm
\begin{abstract}
We use multi-parton states to examine
the leading order collinear factorization of single transverse-spin asymmetries in Drell-Yan processes.
Twist-3 operators are involved in the factorization. We find
that the so-called soft-gluon-pole contribution in the factorization must exist in order to make
the factorization correct. This contribution comes from the corresponding cross-section at
one-loop, while the hard-pole contribution in the factorization comes from the cross-section at
tree-level. Although the two contributions come from results at different orders,
their perturbative coefficient functions in the factorization
are at the same order. This is in contrast to factorizations only involving twist-2 operators.
The soft-gluon-pole contribution found in this work is in agreement with that derived in a different way.
For the hard-pole contributions we find an extra contribution from an extra parton process contributing
to the asymmetries. We also solve a part of discrepancy in evolutions of the twist-3 operator.
The method presented here for analyzing the factorization can be generalized to other processes and can be easily
used for studying factorizations at higher orders, because the involved calculations are of standard
scattering amplitudes.

\vskip 5mm
\noindent
\end{abstract}
\vskip 1cm

\par\noindent
{\bf 1. Introduction}
\par\vskip10pt
Single transverse-spin asymmetries(SSA) have been observed in various experiments, where an involved hadron
is transversely polarized.
A review about the phenomenologies  of SSA can be found in \cite{Review}. In general SSA
can be generated if scattering amplitudes have
nonzero absorptive parts and there are helicity-flip interactions.
It has been shown in production of heavy quarks like
top quarks there are sizeable SSA\cite{KPR,DGBB}. Because a top quark is heavy, the bound state effects
can be neglected. The helicity-flip is due to the nonzero quark mass. Therefore,
in the studies of \cite{KPR,DGBB} one essentially deals with point-like
particles and can use the perturbative theory in standard way.
\par
For SSA involving light hadrons,
the origin of SSA is unclear because of bound-sate effects and the helicity-conservation
of QCD with light quarks which can approximately be taken as massless.
However, certain predictions can be made for SSA in cases where large momentum transfers
are involved. In these cases, one can use the concept of factorization in QCD.
SSA can be factorized in the form of a convolution with nonperturbative matrix elements of hadrons and
perturbative coefficient functions. With this form of predictions one is then able
to explore hadron structures with experiment.
The coefficient functions can be calculated as a perturbative expansion.
In this work we study the collinear factorization of SSA in Drell-Yan processes. We will show
that the factorization of collinear divergences corresponding to nonperturbative effects
is not made in an usual way as factorizations at leading twists.
\par
The collinear factorization for describing SSA has been proposed in \cite{QiuSt, EFTE}.
With the collinear factorizations SSA in various processes has been studied
in \cite{KaKo,EKT,tw31,tw32,JQVY1,JQVY2,KVY}.
In such a factorization, the nonperturbative effects of the transversely polarized hadron
are factorized into twist-3 matrix elements,
or called ETQS matrix elements. Taking SSA in Drell-Yan processes as an example,
SSA is factorized as a convolution of three parts: The first part is the standard parton distribution
function of the unpolarized hadron defined with twist-2 operators. The second part consists of
matrix elements of the polarized hadron defined with twist-3 operators.
The third part consists of perturbative coefficient functions.
The differential cross-section is determined by those twist-2- and twist-3 matrix elements
of the initial hadrons and a forward hard scattering of partons from the initial hadrons.
The perturbative coefficient functions describe the forward hard scattering.
If the factorization can be proven, the coefficient
functions can be calculated safely as a perturbative expansion and they are free from any soft divergence
like collinear- and I.R. divergence.
In this approach the effects of helicity-flip
are parameterized with twist-3 matrix elements, while the absorptive part is generated
in the hard scattering of partons.
\par
The above mentioned collinear factorization has been derived in a
rather formal way by using diagram expansion at hadron level, in
which one divides diagrams into three parts. Two of them are related
to the initial hadrons respectively, the remaining one is related to
the parton scattering. By expanding the two parts related to hadrons
according to twist of operators and the part for the parton
scattering with large momentum transfers, one obtains the factorized
form of SSA and the perturbative coefficient function at leading
order of $\alpha_s$. The forward hard scattering is participated by
three partons from the polarized hadron, i.e., two quarks with one
gluon and two partons from the unpolarized hadron, e.g., $\bar q +
(q+g)\to \gamma^* +X \to \bar q +q$ with the antiquark from the
unpolarized hadron and other partons from the polarized one. It
seems difficult with this method to derive the perturbative
coefficient function at higher orders and to prove the
factorization. It is interesting to note that the contributions in
the factorization consist of two parts. One part of the
contributions is with the gluon carrying a nonzero momentum, called
as hard-pole contributions, while another part of the contributions
is with the gluon carrying zero momentum, called as soft-gluon-pole
contributions. The two parts are associated  with perturbative
coefficient functions starting at the same leading order of
$\alpha_s$.
\par
It should be noted that QCD
factorizations, if they are proven, are general properties of QCD.
These factorizations hold not only with hadron states but also hold when one replaces
the hadron states with parton states. This is in the sense that
the perturbatively calculable parts in factorizations do not depend
on hadrons and  are completely determined by hard scattering of
partons. The procedure for this in the case of SSA is the following:
With partonic states one can calculate the differential cross-section related to SSA
and those twist-2 and twist-3 matrix elements
with perturbative theory. In general they will contain soft divergences which usually appear
beyond leading order. By writing the differential cross-section as a convolution
of these matrix elements with a perturbative coefficient function, one can determine
the function. Beyond the leading order of $\alpha_s$, one may be able
to show that the function is free from soft divergences. If it is true, then the factorization for SSA
is proven. This procedure also provides a way to determine the higher-order corrections to the perturbative
coefficient function.
\par
In our previous works\cite{MS1,MS2,MS3} we have made such an attempt
to derive the factorization by replacing hadrons with partons. To have helicity-flip
with massless partons we have constructed a multi-parton state to replace
the transversely polarized hadron in \cite{MS3}. But with our partonic results at leading
order of $\alpha_s$ we can only find the hard-pole contributions.
This is apparently in contradiction with the early results.
\par
In factorizations only with leading twist-2 operators, it is interesting to note
the following fact:
For a differential cross-section the factorizations at leading order
is completely determined by partonic results at tree-level, i.e.,
the perturbative coefficient functions
at leading order are determined only with the differential cross-section
and twist-2 matrix elements calculated at tree-level with parton states.
This has the implication for one-loop results. If the factorization is right or proven, then
the collinearly divergent part of the differential cross-section with parton states at one-loop
is completely determined by the convolutions of the leading-order perturbative coefficient functions
with the one-loop matrix elements of twist-2 operators.
This patten about the collinearly divergent part of the differential cross-section
can be iterated beyond one-loop.
Assuming this is also the case for the factorization involving twist-3 operators,
one then expects that
the collinearly divergent part of the differential cross-section related to SSA
at one-loop
is completely determined by the convolution of the collinearly divergent parts of twist-2 and twist-3
matrices at one-loop with the perturbative coefficient functions at leading order. However,
the assumptions may not  be correct. Factorizations involving twist-3 operators can be different
than those only with twist-2 operators.

\par
In order to clarify this issue we go beyond the leading order in this work.
We find that certain contributions at one-loop,
which contain collinear divergences, can not be factorized in the way as expected in the above.
To make the factorization with the partonic states correct, one has to introduce  additional contributions
in the derived
factorization which only contains hard-pole contributions. These additional
contributions are just the soft-gluon contributions. This is an interesting fact
because a part of the leading-order perturbative coefficient function
is determined by quantities at non-leading order. It will be
important for  calculating higher-order corrections and  proving the factorization, following
the above outlined procedure. In this work we restrict ourself to the case where
the forward hard scattering of those partons consisting of two antiquark
from the unpolarized hadron and two quarks with one gluon from the polarized hadron. Beside
the above mentioned contributions we also find an extra hard-pole contribution
corresponding to the forward hard scattering where only the gluon  is in the initial- or final state.
By calculating the twist-3 matrix element corresponding to this hard scattering, we can solve
a part of discrepancies in the evolution of the matrix element studied in \cite{KQ,BMP,ZYL}.
\par
In this work we study SSA in Drell-Yan processes in the kinematic
limit of the small transverse momentum of the observed lepton pair.
In this kinematic region there exists another factorization for SSA,
called Transverse-Momentum-Dependent(TMD) factorization, similarly
to the TMD factorization for unpolarized cases studied in
\cite{CS,CSS,JMY,JMYG,CAM}. In the polarized case the
nonperturbative effects of the polarized hadron are factorized into
Sivers function\cite{Sivers} which contains both helicity-flip- and
T-odd effects. The properties of Sivers function  and SSA with it
have been studied extensively
\cite{JC,SJ1,TMDJi,Mulders97,Boer03,Anselmino,Mulders,DeSanctis,Efremov,BQMa}.
It should be noted that in the kinematic region of the small
transverse momentum limit two factorizations apply, if the
transverse momentum is much larger than the QCD scale
$\Lambda_{QCD}$. It has been shown that the two factorizations are
equivalent in the region\cite{JQVY1,JQVY2,KVY}. Again, the TMD
factorization here is derived in the formal way by using the
mentioned diagram expansion at hadron level. In \cite{MS1,MS2,MS3}
we have examined the TMD factorization of SSA  with parton states
and found an agreement with existing results. In this work we will
only focus on the collinear factorization.
\par
Our work is organized as the following: In Sect.2 we give our notations
for Drell-Yan processes and the factorization of SSA. In Sect.3 we introduce
our multi-parton states and give some relevant results of twist-3 matrix elements.
In Sect.4 we calculate SSA with our multi-parton state from certain classes of one-loop
contributions which are collinearly divergent.
We then show that these contributions can not be factorized as the usual way discussed in the above.
These contributions in fact have to be identified as the mentioned soft-gluon-pole contributions
in order to have finite corrections at higher orders of perturbative coefficient functions.
In Sect. 5 we study the extra contribution which should be added to the factorization formula.
There we also show that a part of the discrepancy in the evolution of the twist-3 matrix element
derived in \cite{KQ,BMP} is solved. Sect.6 is our summary and outlook.

\par\vskip20pt
\noindent
{\bf 2. Collinear Factorization of SSA in Drell-Yan Processes}
\par\vskip10pt
We will use the  light-cone coordinate system, in which a
vector $a^\mu$ is expressed as $a^\mu = (a^+, a^-, \vec a_\perp) =
((a^0+a^3)/\sqrt{2}, (a^0-a^3)/\sqrt{2}, a^1, a^2)$ and $a_\perp^2
=(a^1)^2+(a^2)^2$. Other notations are:
\begin{equation}
  g_\perp^{\mu\nu} = g^{\mu\nu} - n^\mu l^\nu - n^\nu l^\mu,
  \ \ \ \ \ \
  \epsilon_\perp^{\mu\nu} =\epsilon^{\alpha\beta\mu\nu}l_\alpha n_\beta, \ \ \ \
  \epsilon^{\alpha\beta\mu\nu} = -\epsilon_{\alpha\beta\mu\nu},  \ \ \ \ \
  \epsilon^{0123}=1
\end{equation}
with the light-cone vectors $l$ and $n$ defined as $l^\mu=(1,0,0,0)$ and $n^\mu=(0,1,0,0)$, respectively.
We consider the Drell-Yan process:
\begin{equation}
  h_A ( P_A, s) + h_B(P_B) \to \gamma^* (q) +X \to  \ell^-  + \ell ^+  + X,
\end{equation}
where $h_A$ is a spin-1/2 hadron with the spin-vector $s$.
We take a light-cone coordinate system in which the momenta and the spin are :
\begin{equation}
P_{A,B}^\mu = (P_{A,B}^+, P_{A, B}^-, 0,0),  \ \ \ \  s^\mu =(0,0, \vec s_\perp).
\end{equation}
$h_A$ moves in the $z$-direction, i.e., $P_A^+$ is the large component. $h_B$ moves
in the $-z$-direction with $P_B^-$ as the large component.  The spin of $h_B$
is averaged. The invariant mass of the observed lepton pair is $Q^2 =q^2$.
The relevant hadronic tensor is defined as a matrix element
of the forward scattering $h_A+ h_B \to \gamma^* +X \to h_A +h_B$:
\begin{equation}
W^{\mu\nu}  = \sum_X \int \frac{d^4 x}{(2\pi)^4} e^{iq \cdot x} \langle h_A (P_A, s_\perp), h_B(P_B)  \vert
    \bar q(0) \gamma^\nu q(0) \vert X\rangle \langle X \vert \bar q(x) \gamma^\mu q(x) \vert
     h_B(P_B),h_A (P_A, s_\perp)  \rangle,
\end{equation}
and the differential cross-section is determined by the hadronic tensor as:
\begin{equation}
\frac{ d\sigma }{ dQ^2 d^2 q_\perp d q^+ d q^- } = \frac{4\pi \alpha_{em}^2 Q_q^2}{3 S Q^2}
    \delta (q^2 -Q^2)
    \left ( \frac {q_\mu q_\nu} {q^2} - g_{\mu\nu} \right ) W^{\mu\nu}, \ \ \  S=2P_A^+ P_B^-.
\end{equation}
\par
We are interested in the kinematical region where $q_\perp^2  \ll  Q^2$. The hadronic tensor
at leading twist accuracy has the structure:
\begin{eqnarray}
W^{\mu\nu} &=& - g_\perp^{\mu\nu} W_U^{(1)} + \left ( g_\perp^{\mu\nu}
-2 \frac{q_\perp^\mu q_\perp^\nu} {q_\perp^2}  \right )
   W_U^{(2)}
\nonumber\\
&&  - g_\perp^{\mu\nu} \epsilon_\perp^{\alpha \beta} s_{\perp\alpha} q_{\perp\beta} W_T^{(1)}
 + \left ( s_{\perp\alpha} \epsilon_\perp^{\alpha\mu} q_\perp^\nu
          +s_{\perp\alpha} \epsilon_\perp^{\alpha\nu} q_\perp^\mu -g_\perp^{\mu\nu}
             \epsilon_\perp^{\alpha \beta} s_{\perp\alpha}  q_{\perp\beta} \right ) W_T^{(2)}
\nonumber\\
  &&  +  q_{\perp\alpha} \left ( \epsilon_\perp^{\alpha\mu}  q_{\perp}^\nu
 + \epsilon_\perp^{\alpha\nu} q_{\perp}^\mu \right ) {\vec  q}_\perp \cdot \vec s_\perp W_T^{(3)}
 +\cdots
\end{eqnarray}
In the above, we only give the structures symmetric in $\mu\nu$. $W_T^{(i)}(i=1,2,3)$ represent
$T$-odd effect related to the spin. $W_U^{(1,2)}$ are responsible for unpolarized cross-sections.
$W_T^{(1)}$ contributes to SSA in the region $q^2_\perp  \ll  Q^2$ which we will study.
We introduce $q^+ =x P^+_A$ and $q^- = y P^-_B$. All structure functions depend on $x$, $y$ and $q^2_\perp$.
\par
In the limit $q_\perp\to 0$ only the structure function $W_T^{(1)}$ gives the leading spin-dependent
contribution to the differential cross-section and hence to SSA. The factorization of the structure function
is the main subject to be studied in this work. In the collinear factorization $W_T^{(1)}$
is factorized with standard parton distributions of hadron $h_B$ and  twist-3 matrix elements
of hadron $h_A$. There are two relevant twist-3 matrix elements. They are defined as:
\begin{eqnarray}
T_F (x_1,x_2)  \epsilon_\perp^{\mu\nu}
s_{\perp\nu}
   & =&    \frac{g_s}{2}\int \frac{dy_1 dy_2}{4\pi}
   e^{ -iy_2 (x_2-x_1) P^+ -i y_1 x_1 P^+ }
\nonumber\\
    && \cdot \left  \{ \langle P, \vec s_\perp \vert
           \bar\psi (y_1n ) \gamma^+ G^{+\mu}(y_2n) \psi(0) \vert P,\vec s_\perp \rangle
  - ( \vec s_\perp \to - \vec s_\perp ) \right \},
\nonumber\\
T_{\Delta,F} (x_1,x_2)
s_{\perp}^{\mu}
   & =&    -i\frac{g_s}{2}\int \frac{dy_1 dy_2}{4\pi}
   e^{ -iy_2 (x_2-x_1) P^+ -i y_1 x_1 P^+ }
\nonumber\\
    && \cdot \left  \{ \langle P, \vec s_\perp \vert
           \bar\psi (y_1n ) \gamma^+ \gamma_5 G^{+\mu}(y_2n) \psi(0) \vert P,\vec s_\perp \rangle
  - ( \vec s_\perp \to - \vec s_\perp ) \right \}.
\label{tw3}
\end{eqnarray}
In the above we have suppressed the gauge links along direction $n$ between operators. These gauge links
make the definition gauge invariant.
The general properties of these twist-3 matrix elements have been discussed in \cite{QiuSt,EKT}. One can show
$T_{\Delta,F}(x,x)=0$. In general $T_F(x,x)$ is not zero. This corresponds to the so-called
soft-gluon-pole contributions because the gluon field in $T_F(x_1,x_2)$ with $x_1=x_2=x$ carries
zero momentum entering the hard scattering.
\par
For the case that SSA is generated through the scattering where an antiquark $\bar q$
from the unpolarized hadron $h_B$ and a quark $q$ or a quark with a gluon from the polarized
hadron $h_A$, i.e, the forward parton scattering $\bar q + q+g \to \gamma^* +X \to \bar q + q$ or the reversed,
 the structure function in the limit
$q_\perp/Q \ll 1$ can be factorized
in the form\cite{JQVY1}:
\begin{eqnarray}
W_T^{(1)}(x,y,q_\perp) &=&
  \frac {\alpha_s}{  (2 \pi q^2_\perp)^2} \int_x^1 \frac{dy_1}{y_1}\int_y^1 \frac{ d y_2}{ y_2} \bar q(y_2)
    \biggr [ {\mathcal A}_h (x,y_1) +{\mathcal A}_s (x,y_1) + {\mathcal A}_{sq}(x,y_1) \biggr ],
\nonumber\\
{\mathcal A}_h (x,y_1) &=&   \delta (1-\xi_2) \left [ T_F(x,y_1) \frac{1+\xi_1}{(1-\xi_1)_+}
    + T_{\Delta, F}(x,y_1) \right ] ,
\nonumber\\
 {\mathcal A}_s (x,y_1) &=& \frac{1}{N_c^2} \left[  y_1 \frac{\partial T_F(y_1,y_1)}{\partial y_1} (1+\xi_1^2) \delta(1-\xi_2)
 +T_F(y_1,y_1) \left ( \frac{2\xi_1^3-3\xi_1^2-1}{(1-\xi_1)_+} \delta(1-\xi_2)
\right.\right.
\nonumber\\
   && \left.\left.   -\frac{\xi_2 (1+\xi_2^2)}{(1-\xi_2)_+} \delta(1-\xi_1) +2 \delta(1-\xi_1)\delta(1-\xi_2) \ln \frac{q^2_\perp}{Q^2}
     \right ) \right ],
\nonumber\\
{\mathcal A}_{sq} (x,y_1) &=& T_F(y_1,y_1) \delta(1-\xi_1)  \left [ \left ( 1 +\frac{\xi_2 -1}{N_c^2} \right) \frac{1+\xi_2^2}{(1-\xi_2)_+}
    -2\delta(1-\xi_2) \ln\frac{q^2_\perp}{Q^2} \right ]   ,
\nonumber\\
     \xi_1 &=& x/y_1, \ \ \  \xi_2=y/y_2.
\label{FAC}
\end{eqnarray}
In the above $\bar q(y_2)$ is the antiquark distribution function of $h_B$.
The contributions to $W_T^{(1)}$ can be divided into three parts.
The part with ${\mathcal A}_h$ consists of the hard-pole contributions.
In  ${\mathcal A}_h$,
the first term with $T_F(x_1,x_2)$ has been first derived in \cite{JQVY1} which is also confirmed
in \cite{MS3}, while the second term with
$T_{\Delta, F}$ has been derived in \cite{MS3}.
The part with  ${\mathcal A}_s$ and that with  ${\mathcal A}_{sq}$  are of the soft-gluon-pole contributions,
because they are related to $T_F(x,x)$.
The soft-gluon-pole contributions in  ${\mathcal A}_{sq}$ only appear in the limit
in the limit $Q\gg q_\perp$.

\par\vskip20pt
\noindent
{\bf 3. Multi-parton State and Twist-3 Matrix elements}
\par\vskip10pt
In \cite{MS1,MS2} we have studied SSA with single-parton states, where the helicity flip is caused
be a finite quark mass $m$. In fact one can study SSA at parton level by taking massless limit $m=0$.
For this one needs to take the effect of helicity flip as that of a correlation between
the spin of quarks and the spin of gluons. For this purpose we consider the state or the system
with the total helicity $\lambda$:
\begin{equation}
 \vert n [\lambda ] \rangle  =  \vert q(p,\lambda_q) [\lambda ] \rangle + c_1
                   \vert q(p_1,\lambda_q) g(k,\lambda_g ) [\lambda ] \rangle,
\label{pas}
\end{equation}
with $p_1+k =p$. In the first term $\lambda_q =\lambda$.
For the $qg$-state, the total helicity is the sum $\lambda_q + \lambda_g$.
We specify the momentum as:
\begin{equation}
   p^\mu =(p^+,0,0,0), \ \ \  p_1^\mu = x_0 p^\mu, \ \ \ \ k^\mu =(1-x_0) p_\mu =\bar x_0 p^\mu.
\end{equation}
The $q$-state and $qg$-state carries the same color index $i_c$ as given:
\begin{equation}
\vert  q (p,\lambda_q) \rangle = b^\dagger_{i_c} (p,\lambda_q) \vert 0 \rangle,
\ \ \ \ \
\vert q (p_1,\lambda_q)g(k,\lambda_g) \rangle =
T^a_{j_c i_c} b^\dagger_{j_c} (p_1,\lambda_q) a^\dagger_a (k,\lambda_g) \vert 0 \rangle,
\end{equation}
where $b^\dagger_i$ is the quark creation operator with $i$ as the color index,
$a^\dagger_a$ is the gluon creation operator with $a$ as the color index.
$c_1$ is taken as a real number.
\par
From standard text book we know that the transverse-spin dependent part
of a matrix element, like the twist-3 matrix elements or the transverse-spin dependent part of $W^{\mu\nu}$,
corresponds to the off-diagonal part of the matrix element in helicity space. Because of helicity conservation
in QCD with massless quarks, the twist-3 matrix elements are always zero if we replace
the hadron in  the matrix elements with a single quark.
However, if one replaces the hadron with the above multi-parton state, the twist-3 matrix elements
receive nonzero contributions from the interference between the single quark state and the state
consisting of a quark and a gluon. In the interference, the quark always has the same helicity, while
the helicity change is due to the helicity of the gluon. The structure function $W_T^{(1)}$
also receives nonzero contributions from the interference.
\par
By replacing the hadron with the multi-parton state one can calculate those twist-3 matrix elements
and the structure functions perturbatively for the purpose of factorization.
At tree-level, it is straightforward to obtain the twist-3 matrix elements as:
\begin{eqnarray}
T_F(x_1,x_2) &=& c_1 g_s\pi \sqrt{\frac{x_0}{2}} (N_c^2-1)(x_2-x_1)  \left [
                      \delta (1-x_1) \delta (x_2-x_0) - \delta (1-x_2) \delta (x_1-x_0) \right ],
\nonumber\\
T_{\Delta,F}(x_1,x_2) &=&  c_1 g_s\pi \sqrt{\frac{x_0}{2}} (N_c^2-1)(x_2-x_1)  \left [
                      \delta (1-x_1) \delta (x_2-x_0) + \delta (1-x_2) \delta (x_1-x_0) \right ].
\end{eqnarray}
These functions are proportional to $c_1$ indicating that they are from the mentioned interference.
For simplicity we will set $c_1=1$ in the following sections without confusion.
\par
\par
\begin{figure}[hbt]
\begin{center}
\includegraphics[width=4cm]{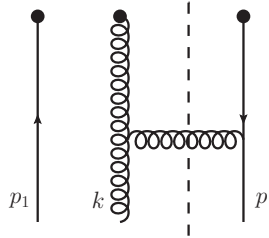}
\end{center}
\caption{A diagram for the twist-3 matrix element $T_F(x_1,x_2)$ which gives
a nonzero contribution with $x_1=x_2=x$. Black dots denote insertion of operators in the definition
of $T_F(x_1,x_2)$. The broken line is a cut.}
\label{Feynman-dg3}
\end{figure}
\par
It is noted that at tree-level the function $T_F(x,x)$ becomes zero. This is the reason
why we can not re-produce in \cite{MS3} the soft-gluon-pole contributions with tree-level results of $T_F$
and $W_T^{(1)}$.
However, the function becomes nonzero at one-loop. To show this, we examine a particular contribution from a one-loop
diagram given in Fig.1 in Feynman gauge. The contributions contain a U.V. divergence and a collinear divergence.
Regularizing these divergences and subtracting the U.V. one we have:
\begin{eqnarray}
T_F(x_1,x_2)\biggr\vert_{Fig.1} &=& \frac{g_s\alpha_s}{16} N_c (N_c^2-1) \sqrt{2x_0}
 \left [ \delta(x_2-x_0)
    \frac{1-x_1}{1-x_0} \left ( \frac{x_0-x_1}{1-x_1} -2 x_0 \right )
\right.
\nonumber\\
  && \left.  +\delta(x_1-x_0)
    \frac{1-x_2}{1-x_0} \left ( \frac{x_0-x_2}{1-x_2} -2 x_0 \right ) \right ]
    \left (-\frac{2}{\epsilon_c} +\gamma -\ln\frac{\mu^2}{4\pi\mu_c^2} \right ),
\nonumber\\
T_{\Delta,F}(x_1,x_2)\biggr\vert_{Fig.1} &=& \frac{g_s\alpha_s}{16} N_c (N_c^2-1) \sqrt{2x_0}
 \left [ \delta(x_2-x_0)
    \frac{1-x_1}{1-x_0} \left ( \frac{x_0-x_1}{1-x_1} -2 x_0 \right )
\right.
\nonumber\\
  && \left.  -\delta(x_1-x_0)
    \frac{1-x_2}{1-x_0} \left ( \frac{x_0-x_2}{1-x_2} -2 x_0 \right ) \right ]
    \left (-\frac{2}{\epsilon_c} +\gamma -\ln\frac{\mu^2}{4\pi\mu_c^2} \right ),
\end{eqnarray}
where the pole in $\epsilon_c =4-d$ represents the collinear divergence and $\mu_c$ is the scale
associated with it. The scale $\mu$ is associated with the subtracted U.V. divergence.
The collinear divergence appears because the gluon going through the cut can be collinear
to the incoming gluon and the outgoing quark in Fig.1. In the above we also give the contribution
to $T_{\Delta,F}$ from Fig.1.
From this result we see that $T_F$ is nonzero at $x_1=x_2$. After examining all one-loop
diagrams in Feynman gauge, we find that only the diagram in Fig.1 gives nonzero contribution
at $x_1=x_2$, i.e.,
\begin{equation}
T_F(x,x) =- \frac{g_s\alpha_s}{4} N_c (N_c^2-1) x_0 \sqrt{2x_0} \delta (x_0-x)
 \left (-\frac{2}{\epsilon_c}  +\gamma -\ln\frac{\mu^2}{4\pi\mu_c^2} \right ).
\label{TFxx}
\end{equation}
This result is in agreement with our previous calculation in the light-cone gauge $n\cdot G=0$\cite{MS3}.
The function $T_{\Delta,F}$ is always zero at $x_1=x_2=x$.
\par
\begin{figure}[hbt]
\begin{center}
\includegraphics[width=6cm]{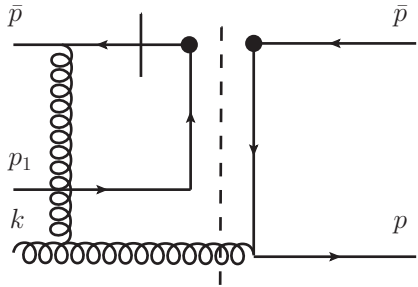}
\end{center}
\caption{The diagram for the tree-level contribution to SSA in the limit $q_\perp\ll Q$.
The black dots represent the insertion of electro-magnetic current operators. The broken
line is the cut.  The short bar cutting
the quark propagator means to take the absorptive part of the propagator.    }
\label{Feynman-dg2}
\end{figure}
\par
To calculate the hadronic tensor, we replace the polarized hadron $h_A$ with the parton state $\vert n \rangle$
of Eq.(\ref{pas}), and the unpolarized hadron $h_B$ with an antiquark with the momentum $\bar p^\mu=(0,\bar p^-,0,0)$.
At tree-level, $W_T^{(1)}$ also receives contributions from the interference. i.e., from the forward
scattering $\bar q + q+g \to \gamma^* + g \to \bar q +q$ and $\bar q + q \to \gamma^* + g \to \bar q +q +g $.
In the limit
$q_\perp\ll Q$, only one diagram given in Fig.2 gives the contribution.
In the diagram the short bar means to take the absorptive part of the cut propagator with the momentum
$k_q$, i.e.,
\begin{equation}
{\rm Abs }\left [ \frac{i\gamma\cdot k_q}{k_q^2 + i\varepsilon} \right ] =  i\gamma\cdot k_q \left (-i\pi \delta (k_q^2) \right ).
\end{equation}
In fact, the short bar
here represents a physical cut of the amplitude of the left part in the diagram.
It has been shown
in \cite{MS3} with the tree-level results of the twist-3 matrix elements given in Eq.(12) that
the tree-level result of $W_T^{(1)}$ only produces the ${\mathcal A}_h$-term  in
Eq.(\ref{FAC}). If one expects that the factorization involving twist-3 operators here
happens in the same way as in the factorization only with twist-2 operators, one will conclude
that at leading order of $\alpha_s$ of the perturbative coefficient function $W_T^{(1)}$
is predicted only with the hard-pole contributions of Eq.({\ref{FAC}). This is obvious in contradiction
with the results in Eq.({\ref{FAC}).

\par\vskip20pt
\noindent
{\bf 4. Soft-Gluon-Pole Contributions}
\par\vskip10pt
\par
At tree-level, all momenta carried by gluon lines in Fig.2 are fixed and can not be zero, e.g.,
the momentum $k_1$ of the gluon crossing the cut
is fixed by the total momentum conservation.
Therefore, we can not identify any gluon line in Fig.2 corresponding to the gluon field
with zero momentum
in $T_F(x,x)$ in Eq.(7).
Now we consider the case in which there is an extra gluon exchanged and crossing the cut
as those diagrams given in Fig.3. In this case the momentum $k_1$ has to be integrated.
In the integration the collinear region, in which the gluon with $k_1$
is collinear to the incoming gluon and outgoing quark, is included. This will result in a collinear divergence.
Taking Fig.3a as such an example, there is an extra gluon exchanged in comparison with Fig.2.
The momentum $k_1$ hence will be integrated. In the collinear region of $k_1$,
one can realize that the part of Fig.3a including
the three gluon vertex and the vertex absorbing the gluon with $k_1$ is essentially given
by Fig.1. This indicates that the collinearly divergent part from the collinear region
may be factorized with $T_F(x_1,x_2)$ or $T_{\Delta,F}$ given in Fig.1 in the form of a convolution of
$T_F$ or $T_{\Delta,F}$  with a perturbative coefficient function.
If the function is not the same
as those in ${\mathcal A}_h$ of Eq.(\ref{FAC}) determined at leading order, then one has to add
extra terms beside the term with ${\mathcal A}_h$  in  Eq.(\ref{FAC}) to make sure that
the factorization is correct at one-loop level.
It is interesting to note that
by taking the absorptive part of the quark propagator in the left part of Fig.3a, one finds
that in the collinear region of $k_1$
the momentum of the gluon exchanged between the initial gluon and initial $\bar q$
is a soft gluon. More precisely, the exchanged gluon is a Glauber gluon with the momentum patten
$k^\mu \sim (\lambda_0^2,\lambda_0^2,\lambda_0,\lambda_0)$ with $\lambda_0 \ll 1$.
This indicates that the possible extra terms may be soft-gluon-pole contributions factorized
with $T_F(x,x)$ from Fig.1.
In this section
we show that this is indeed the case.
\par
\begin{figure}[hbt]
\begin{center}
\includegraphics[width=11cm]{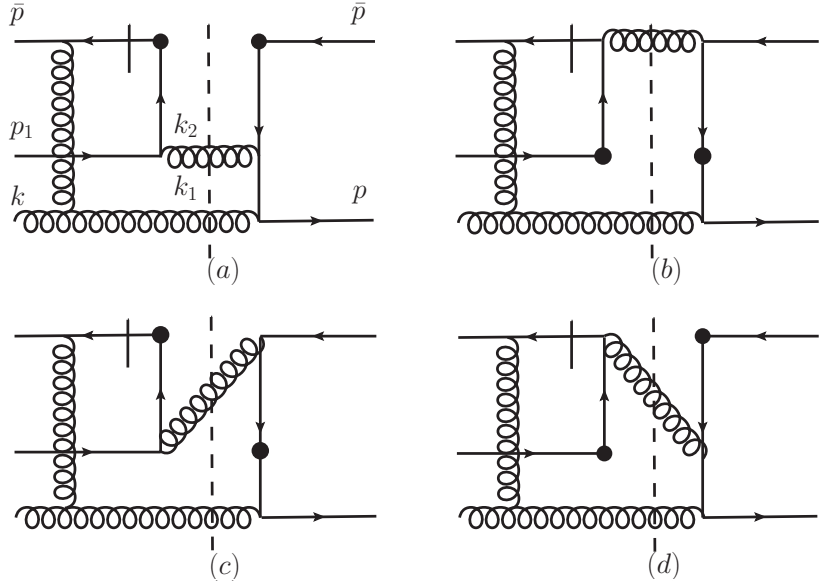}
\end{center}
\caption{The diagrams for the amplitude
$\bar q + q + G \to \gamma^* + G + G \to \bar q + q$, which gives a part of contributions
to SSA. } \label{Feynman-dg5}
\end{figure}
\par
As discussed in the above, it is easy to find those diagrams at one-loop which can give contributions
to the soft-gluon-pole contributions.
The diagrams are those where the incoming gluon emits a collinear gluon and the collinear gluon
is absorbed by an outgoing collinear quark. Beside these gluons, there is an extra exchanged gluon crossing
the cut responsible for the finite $q_\perp$. A class of those diagrams is given in Fig.3.
For our purpose we consider those contributions from the collinear region
where the gluon crossing the cut and emitted by the initial gluon is collinear
to the initial gluon and the outgoing quark.
\par
We take Fig.3a as an example
to show how we obtain the collinearly divergent part.
The contribution from Fig.3a to the hadronic tensor can be written with standard Feynman rule as:
\begin{eqnarray}
W^{\mu\nu} \biggr\vert_{3a} &=& -\frac{ig_s^5}{2N_c} f^{a_1a_3a} \int \frac{d^4 k_1}{(2\pi)^4}\frac{d^4 k_2}{(2\pi)^4}
 \delta^4 (\bar p + p -k_1-k_2 -q) (2\pi\delta(k_1^2)) (2\pi\delta (k_2^2))(-i\pi \delta ((k_3 +\bar p)^2)
)\nonumber\\
  && \cdot \biggr [ \bar v(\bar p) \gamma^{\mu_3} \gamma\cdot(k_3+\bar p) \gamma^\mu \gamma\cdot (p_1-k_2) \gamma^{\rho_2}
     T^{a_3} T^{a_2} T^a u(p_1,\lambda_q)
\nonumber\\
   &&  \ \cdot \bar u(p,\lambda_q) T^{a_1} T^{a_2} \gamma_{\rho_1} \gamma\cdot (p-k_1) \gamma_{\rho_2} \gamma\cdot (p-k_1-k_2)
   \gamma^\nu v(\bar p)
\nonumber\\
   && \ \left ( \epsilon \cdot (-k_1+k_3) g^{\rho_1}_{\ \ \mu_3} +(-k_3-k)^{\rho_1}\epsilon_{\mu_3}
     +(k_1+k)_{\mu_3} \epsilon^{\rho_1} \right ) \biggr ]
\nonumber\\
   && \cdot \frac{1}{((p-k_1-k_2)^2 - i\varepsilon)((p-k_1)^2 -i\varepsilon)( (p_1-k_2)^2 + i\varepsilon)
      (k_3^2 + i\varepsilon)},
\label{3a}
\end{eqnarray}
the color and spin of the initial antiquark $\bar q$ is averaged and it gives the factor $(2 N_c)^{-1}$.
The initial quark has the helicity $\lambda_q$, the initial gluon has $\lambda_g$.
The absorptive part in the scattering amplitude is generated by the cut cutting the quark propagator.
This gives the $\delta$-function $\delta((k_3+\bar p)^2)$ with $k_3$ being the momentum carried
by the gluon exchanged between the initial gluon and the initial antiquark in the left part of Fig.3a.
We will consider
the collinear region where the momentum $k_1$ of the gluon emitted by the initial gluon
is collinear. In the collinear region the momentum $k_1$ scales as:
\begin{equation}
   k_1^\mu \sim (1,\lambda_0^2,\lambda_0, \lambda_0)
\label{ck1}
\end{equation}
with $\lambda_0\ll 1$.
The on-shell condition from the quark propagator fixes $k_1^+$ in the collinear region as
\begin{equation}
\frac{1}{(k_3+\bar p)^2+ i\varepsilon} \Rightarrow  -i\pi \delta((k_3+\bar p)^2)
\ \Rightarrow k_1^+ \approx k^+ -\frac{k^2_{1\perp}}{2 \bar p^-} + {\mathcal O}(\lambda_0^4).
\end{equation}
This constraint leads to that the gluon with $k_3$
is a Glauber gluon. It is soft
and may be represented by the gluon field in  $T_F(x,x)$.
\par
The evaluation of the contribution containing the collinear divergence from
the collinear region given by Eq.(\ref{ck1}) is rather straightforward. One first uses
these $\delta$-functions to perform the integration over $k_2$, $k_1^-$ and $k_1^+$.
The remaining integration is that over $k_{1\perp}$. The integrand is then besides some trivial factors
a product of those terms in $[ \cdots ]$ in Eq.(\ref{3a}), the denominators of propagators and
the $\delta$-function $\delta (k_2^2)$. Now one can expand the integrand in $\lambda_0$. The leading order is
at $\lambda_0^{-3}$ which does not give the collinear divergence. The next-to-leading order
is at $\lambda_0^{-2}$, which give the collinear divergence after the integration over $k_{1\perp}$.
Contributions from higher orders are finite. In the expansion
we notice that the $\delta$-function $\delta (k_2^2)$ also depends on $k_1$ and needs to be expanded.
The expansion will give a contribution proportional to the derivative of the $\delta$-function. This contribution
may correspond to those terms in Eq.(\ref{FAC}) with the derivative of $T_F(x,x)$.
\par
After the integration over $k_{1\perp}$ one can take the limit $q_\perp\ll Q$. To derive the limit
we will use the following  in the limit of $ q_\perp \to 0$:
\begin{eqnarray}
s \delta(s(1-y)(x_0-x)-q^2_\perp )&\approx & \frac{\delta(1-y)}{(x_0-x)_+} + \frac{\delta(x_0-x)}{(1-y)_+}
  -\delta(x_0-x)\delta(1-y)\ln\frac{q^2_\perp}{Q^2},
\nonumber\\
s \delta'(s(1-y)(x_0-x)-q^2_\perp ) &\approx & \frac{1}{q^2_\perp}\delta(x_0-x)\delta(1-y),
\ \ \ \delta'(u) = \frac{d \delta (u)}{d u},
\end{eqnarray}
with $s=2 p^+ \bar p^-$. The calculation can be simplified by the following: We need
only to calculate the off-diagonal part of the matrix element in helicity space. In this part
one always has $\lambda_q \lambda_g =-1$. We will set $\lambda_q \lambda_g =-1$ in our calculation
for simplicity.
We have the collinearly divergent part of the hadronic tensor in the limit $q_\perp \ll Q$ :
\begin{eqnarray}
W^{\mu\nu} \biggr\vert_{3a} = -i \frac{g_s \alpha_s^2}{(4 \pi)^2}\frac{N_c^2-1}{ N_c}
  g_\perp^{\mu\nu}\frac {\vec\epsilon(\lambda_g) \cdot \vec q_\perp}{(q^2_\perp)^2}
    \sqrt{x_0} \frac{x_0^2-x^2}{x_0^2} \delta (1-y) \left (-\frac{2}{\epsilon_c} \right )
    +\cdots.
\end{eqnarray}
where $\cdots$ stand for the following contributions:  the contributions at non-leading order
in the expansion in $q_\perp/Q$, the contributions which do not contain the collinear
divergence and the contributions of tensor structures other than $g_\perp^{\mu\nu}$. These contributions
are irrelevant for our purpose.
By adding the complex
conjugated contribution with different parton helicities as a part of the interference, one can then obtain the off-diagonal part
of the hadronic tensor in the helicity space.
From the off-diagonal part we can extract the structure function:
\begin{equation}
W_T^{(1)}\biggr\vert_{3a} =\frac{g_s \alpha_s^2}{(4 \pi)^2}\frac{N_c^2-1}{ N_c}
  \frac{1}{(q^2_\perp)^2}
    \sqrt{2 x_0}\frac{x_0^2-x^2}{x_0^2}  \delta (1-y) \left (-\frac{2}{\epsilon_c} \right ) +\cdots,
\end{equation}
again, $\cdots$ denote those irrelevant contributions which do not contain the collinear divergence
or are not at the leading order in the expansion in $q_\perp/Q$.
In the limit the leading order of $W_T^{(1)}$ is of $q^{-4}_\perp$.
In the following we will only give the collinearly divergent contributions
in the limit of $q_\perp\to 0$ explicitly.
Performing similar calculations we have the contributions from other diagrams
in Fig.3:
\begin{eqnarray}
W^{(1)}_T \biggr\vert_{3b} &=& {\mathcal O}(q_\perp^{-2}),
\nonumber\\
W^{(1)}_T \biggr\vert_{3c} &=& \frac{g_s \alpha_s^2}{(4 \pi)^2}\frac{N_c^2-1}{N_c} \left (-\frac{2}{\epsilon_c} \right )
      \frac{\sqrt{2 x_0}}{(q^2_\perp)^2}
  \biggr \{ \frac{x_0 y \delta (x_0-x)}{(1-y)_+} + \frac{x \delta(1-y)}{(x_0-x)_+}
\nonumber\\
   && \ \ \ \ \ \  +x_0 \delta(x_0-x)\delta(1-y) \left ( 1 -\ln\frac{q^2_\perp}{Q^2} \right ) \biggr \}
     +\cdots,
\nonumber\\
W^{(1)}_T \biggr \vert_{3d}  &=& \frac{g_s \alpha_s^2}{(4 \pi)^2}\frac{N_c^2-1}{N_c} \left (-\frac{2}{\epsilon_c} \right )
      \frac{\sqrt{2 x_0}}{(q^2_\perp)^2}
  \biggr \{ \frac{x_0 y^2 \delta (x_0-x)}{(1-y)_+} + \frac{x^2 \delta(1-y)}{x_0 (x_0-x)_+}
\nonumber\\
    &&  \ \ \ \ \  +x_0 \delta(x_0-x)\delta(1-y) \left ( 1 -\ln\frac{q^2_\perp}{Q^2} \right ) \biggr \}
    +\cdots,
\end{eqnarray}
\par

\begin{figure}[hbt]
\begin{center}
\includegraphics[width=11cm]{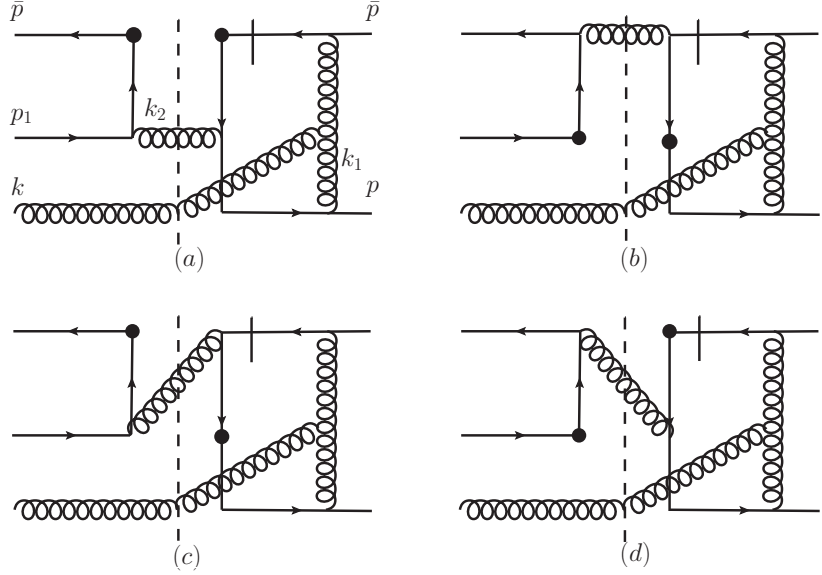}
\end{center}
\caption{The diagrams for the amplitude
$\bar q + q + G \to \gamma^* + G + G \to \bar q + q$, which gives a part of contributions
to SSA.} \label{Feynman-dg5}
\end{figure}
Beside those diagrams in Fig.3, there is another class of diagrams which give the wanted
contributions. The calculations of these diagrams are slightly different than those of Fig.3.
We illustrate this by taking Fig.4a as an example. The contribution from Fig.4a can be written as:
\begin{eqnarray}
W^{\mu\nu} \biggr\vert_{4a} &=&  -\frac{g_s^5}{2N_c} f^{a_1a_3a} \int \frac{d^4 k_1}{(2\pi)^4}\frac{d^4 k_2}{(2\pi)^4}
 \delta^4 (\bar p + p_1 -k_2 -q) (2\pi\delta (k_2^2))(i\pi\delta ((k_3+\bar p)^2)
\nonumber\\
  && \biggr [ \bar u(p)\gamma^{\rho_1} \gamma\cdot (p-k_1) \gamma^{\rho_2}\gamma\cdot (p-k_1-k_2) \gamma^\nu
  \gamma\cdot (\bar p +k_3) \gamma^{\mu_3} T^{a_1}T^{a_2}T^{a_3} v(\bar p)
\nonumber\\
   && \cdot \bar v(\bar p) \gamma^\mu\gamma\cdot (p_1-k_2) \gamma_{\rho_2} T^{a_2} T^a u(p_1)
 \left ( \epsilon \cdot (-k_1-k_3) g_{\rho_1 \mu_3} +(k_3-k)_{\rho_1}\epsilon_{\mu_3}
     +(k_1+k)_{\mu_3} \epsilon_{\rho_1} \right ) \biggr ]
\nonumber\\
   && \cdot \frac{1}{((p-k_1-k_2)^2 - i\varepsilon)((p-k_1)^2 -i\varepsilon)( (p_1-k_2)^2 + i\varepsilon)
      (k_3^2 - i\varepsilon)( k_1^2 -i\varepsilon)}.
\label{W4a}
\end{eqnarray}
Again, the initial quark has the helicity $\lambda_q$, the gluon has $\lambda_g$.
The cut of the quark propagator gives the $\delta$-function $\delta ((k_3+\bar p)^2)$ with $k_3=k_1-k$
as the momentum of the gluon emitted by the antiquark in the right part of Fig.4a.
$k_1$ is the momentum of the gluon emitted by the quark in the right part of Fig.4a.
Unlike the gluon with $k_1$ in Fig.3a, where it is on-shell,
the gluon with $k_1$ in Fig.4a is off-shell in general.
This difference results in that the integration over $k_1^-$ looks nontrivial
at the first step, while the integration over $k_1^-$ in Fig.3a can be simply done
with the on-shell condition $\delta (k_1^2)$.
\par
Now we consider the collinear region where $k_1$ is collinear to $k$ and $p$. The scaling
of its each component is given in Eq.(\ref{ck1}). Then from the $\delta$-function $\delta((k_3+\bar p)^2)$
$k_1^+$ is fixed as $k_1^+ =k^+ +{\mathcal O}(\lambda_0^2)$ after the integration over $k_1^+$.
The integration over $k_1^-$ can be done with a contour in the complex $k^-_1$-plan.
With the fixed $k_1^+$ one can show from Eq.(\ref{W4a}) that there are poles from denominators
of quark propagators only in the lower-half plan. These poles are corresponding physical cuts.
One can use these poles by taking a contour in the lower-half plan to perform the integration.
However,  we notice that there is only one pole in
the upper-half plan. The pole is from the gluon propagator with the momentum $k_1$.
One can equivalently use this pole by taking a contour
in the upper-half plan to perform the integration. Therefore, the integration over $k_1^-$
can be done easily by the replacement:
\begin{equation}
  \frac{1}{k_1^2-i\varepsilon} \to 2\pi i \delta(k_1^2).
\end{equation}
This also applies
for other three diagrams in Fig.4.
It is interesting to note that the gluon with $k_1$ or with $k_3$ in Fig.4a
are in corresponding to the gluon with $k_1$ or with $k_3$ in Fig.3a, respectively.
The gluon with $k_3$ in Fig.4a is also a Glauber gluon.
The remaining calculations are similar
to those of Fig.3. We have the following results of $W_T^{(1)}$ from Fig.4:
\begin{eqnarray}
W^{(1)}_T\vert_{4a} &=& - \frac{g_s\alpha_s^2}{(4\pi)^2} \frac{N_c^2-1}{N_c} \frac{\sqrt{2x_0}}{x_0^2}
  \frac{1}{(q^2_\perp)^2} \left (-\frac{2}{\epsilon_c} \right )
   x(x_0-x)\delta(1-y),
\nonumber\\
W^{(1)}_T\vert_{4b} &=&  \frac{g_s\alpha_s^2}{(4\pi)^2} \frac{N_c^2-1}{N_c} \sqrt{2x_0}
  \frac{1}{(q^2_\perp)^2} \left (-\frac{2}{\epsilon_c} \right )
   x_0 y(1-y)\delta(x_0-x),
\nonumber\\
W^{(1)}_T\vert_{4c} &=& -\frac{g_s\alpha_s^2}{(4\pi)^2} \frac{N_c^2-1}{N_c} \sqrt{2x_0}
  \frac{1}{(q^2_\perp)^2} \left (-\frac{2}{\epsilon_c} \right )
   x_0 y \delta(x_0-x),
\nonumber\\
W^{(1)}_T\vert_{4d} &=& \frac{g_s\alpha_s^2}{(4\pi)^2} \frac{N_c^2-1}{N_c} \frac{\sqrt{2x_0}}{x_0}
  \frac{1}{(q^2_\perp)^2} \left (-\frac{2}{\epsilon_c} \right )
    x\delta(1-y).
\end{eqnarray}
Summing the contributions from Fig.3 and Fig.4 together we obtain the collinearly divergent contribution
of $W_T^{(1)}$, denoted as $W^{(1)}_{T,s}$:
\begin{eqnarray}
W^{(1)}_{T,s}&=& \frac{g_s\alpha_s^2}{(4\pi)^2} \frac{N_c^2-1}{N_c} \left (-\frac{2}{\epsilon_c} \right )
  \frac{\sqrt{2x_0}}{(q_\perp^2)^2}
\left  [ 2 \delta(1-\zeta)\delta(1-y) \left ( 1 -\ln\frac{q^2_\perp}{Q^2} \right )
\right.
\nonumber\\
    && \left. + \delta (1-\zeta) \frac{y(1+y^2)}{(1-y)_+}
       - \delta (1-y) \left ( \frac{2\zeta^3-3\zeta^2-1}{(1-\zeta)_+} + 2 \zeta^2 \right )\right ] ,
\nonumber\\
  \zeta &=& \frac{x}{x_0}.
\end{eqnarray}
\par
If the factorization here takes the same patten as those only with twist-2 operators as
discussed in the introduction and assuming that there is only the hard-pole contribution
at leading order, one then expects that the above $W^{(1)}_{T,s}$ should obey:
\begin{equation}
W^{(1)}_{T,s} =
  \frac {\alpha_s}{  (2 \pi q^2_\perp)^2} \delta (1-y) \int_x^1 \frac{dy_1}{y_1}
   \left [  \frac{1+\xi_1}{(1-\xi_1)_+} T_F(x,y_1)\biggr\vert_{Fig.1}
    + T_{\Delta, F}(x,y_1)\biggr\vert_{Fig.1}  \right ]
\end{equation}
so that the collinear divergences caused by the collinear gluon in Fig.3 and Fig.4 will not appear
in the perturbative coefficient functions at one-loop. In the above we have already taken the tree-level
result $\bar q(y_2)=\delta(1-y_2)$. It is easy to see that the above equation does not hold
because the color factor of $W^{(1)}_{T,s}$ does not match that of $T_F(x,y_1)$ and $T_{\Delta,F}(x,y_1)$ from Fig.1.
Therefore, the factorization at leading order must contain extra terms besides the hard-pole contributions.
\par
With the discussion at the beginning of this section, parts in each diagram in Fig.3 and Fig.4
can be identified with Fig.1. By deleting these parts, these one-loop diagrams
reduce to those for the forward scattering $\bar q + q^* + g^* \to \gamma^* +X \to \bar q + q^*$.
In the kinematic region considered here, the off-shell quark can be approximately taken as an on-shell
quark. The virtual gluon is the mentioned Glauber gluon. It carries the momentum with the $+$-component
approaching to zero. Therefore, $W^{(1)}_{T,s}$ should be factorized with $T_F(x,x)$ from Fig.1. With this observation,
we realize that the factorization of this part is exactly given by the term with ${\mathcal A}_s$ of
Eq.(\ref{FAC}). It is interesting to note that the term with the derivative of $T_F(x,x)$ is also reproduced.
Therefore, we can already conclude here that factorizations involving twist-3 operators are different
than those factorizations only with twist-2 operators. The perturbative coefficient
function in a factorization involving twist-3 operators can not be completely determined by tree-level results
of the differential cross-section and twist-3 matrix elements.

\par
Besides the contributions studied in the above there are soft-gluon-pole
contributions appearing in the limit $q_\perp\to 0$.
There are two classes of diagrams given in Fig.5 and Fig.6. Their calculations are similar
to those in Fig.3 and Fig.4. We only list their divergent contributions:

\par
\begin{figure}[hbt]
\begin{center}
\includegraphics[width=12cm]{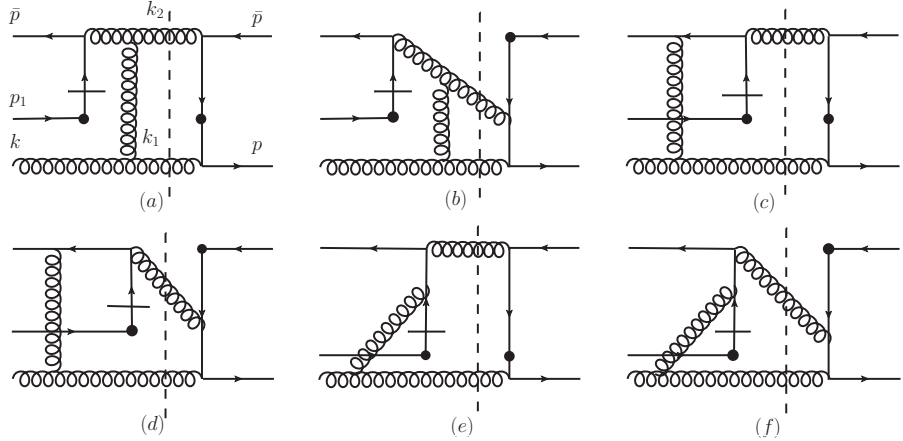}
\end{center}
\caption{The diagrams for soft-gluon-pole contributions appearing in the limit $q_\perp \to 0$}
\end{figure}
\par

\begin{eqnarray}
W^{(1)}_T\vert_{5a} &=&- \frac{g_s \alpha_s^2}{(4\pi)^2} N_c (N_c^2-1)
\frac{ 1 }{(q^2_\perp)^2} \left (-\frac{2}{\epsilon_c}\right )
\sqrt{2 x_0} x_0 \delta (x-x_0) y +\cdots,
\nonumber\\
W_T^{(1)} \vert_{5d}  &=&  -\frac{g_s \alpha_s^2}{(4\pi)^2}\frac {(N_c^2-1)}{N_c} \left (-\frac{2}{\epsilon_c} \right )
\frac{ x_0\sqrt{2 x_0} }{(q^2_\perp)^2} \delta(x-x_0) \left [ \frac{y^2}{(1-y)_+}
    -\delta (1-y) \ln\frac{q^2_\perp}{Q^2} \right ],
\nonumber\\
W_T^{(1)} \vert_{5e}  &=&  -\frac{g_s \alpha_s^2}{(4\pi)^2}\frac {(N_c^2-1)^2}{N_c} \left (-\frac{2}{\epsilon_c} \right )
\frac{ x_0\sqrt{2 x_0} }{(q^2_\perp)^2} \delta(x-x_0) (1-y),
\nonumber\\
W_T^{(1)} \vert_{5f}  &=&  -\frac{g_s \alpha_s^2}{(4\pi)^2}\frac {(N_c^2-1)^2}{N_c} \left (-\frac{2}{\epsilon_c} \right )
\frac{ x_0\sqrt{2 x_0} }{(q^2_\perp)^2} \delta(x-x_0) \left [ \frac{y}{(1-y)_+}
    -\delta (1-y) \ln\frac{q^2_\perp}{Q^2} \right ],
\end{eqnarray}
the contributions from Fig.5b and Fig.5c are non-leading in the limit of $q_\perp/Q \ll 1$.

\par
\begin{figure}[hbt]
\begin{center}
\includegraphics[width=12cm]{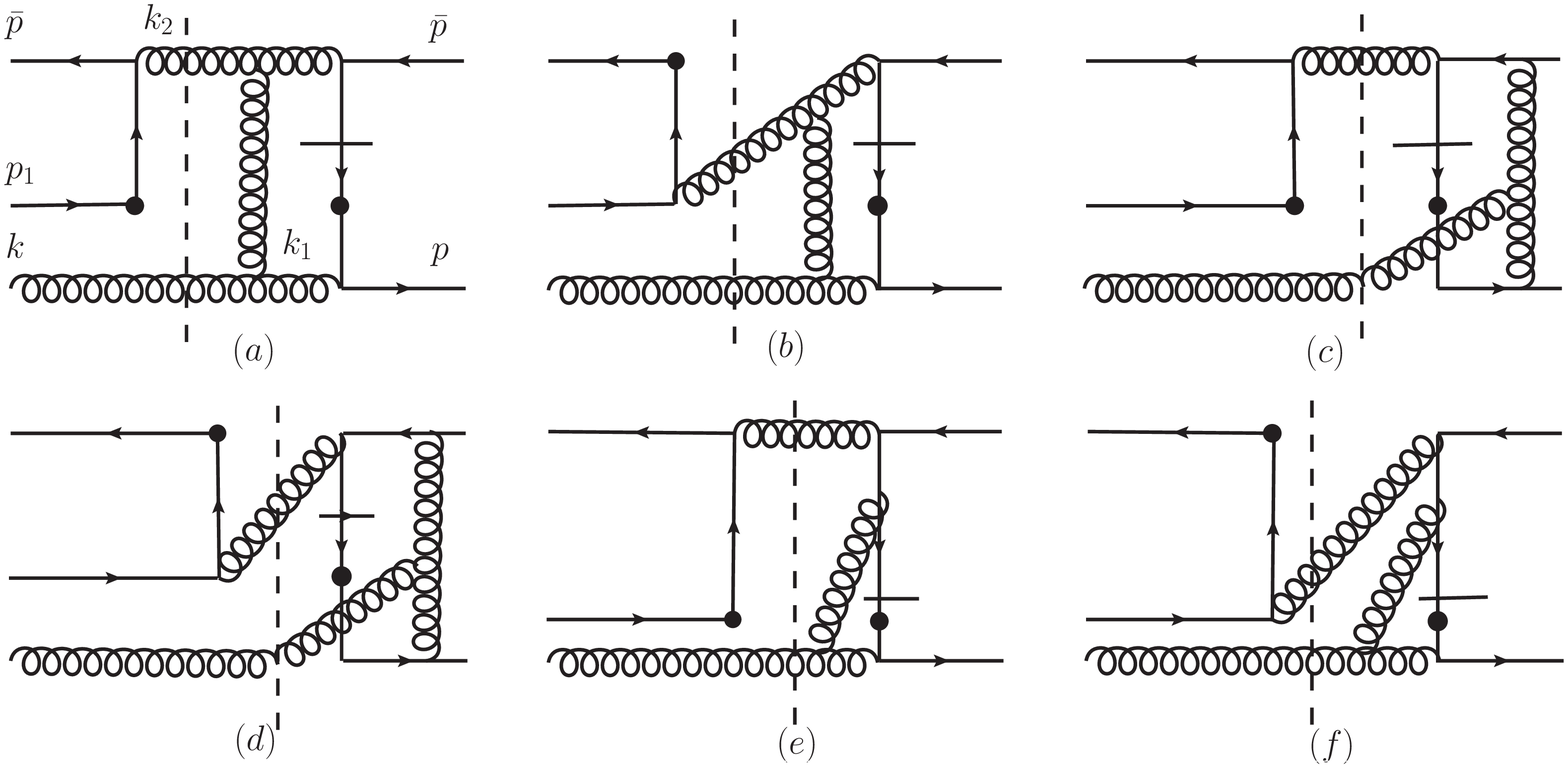}
\end{center}
\caption{The diagrams for soft-gluon-pole contributions appearing in the limit $q_\perp \to 0$}
\end{figure}
\par
The contributions from Fig.6 are:
\begin{eqnarray}
W_T^{(1)}\biggr\vert_{6a} &=& -\frac{g_s\alpha_s^2}{(4\pi)^2} \frac{N_c(N_c^2-1)}{2} \frac{\sqrt{2 x_0}}{(q^2_\perp)^2}
  \left ( -\frac{2}{\epsilon_c}\right ) \biggr [ \frac{\delta(x_0-x)}{(1-y)_+} x_0 y (2y-1)
\nonumber\\
   &&  \ \ \ +\frac{\delta (1-y)}{(x_0-x)_+} \frac{x^2}{x_0}- x_0 \delta(x_0-x)\delta (1-y) \ln\frac{q^2_\perp}{Q^2} \biggr ],
\nonumber\\
W_T^{(1)}\biggr\vert_{6b} &=& -\frac{g_s\alpha_s^2}{(4\pi)^2} \frac{N_c(N_c^2-1)}{2} \frac{\sqrt{2 x_0}}{(q^2_\perp)^2}
  \left ( -\frac{2}{\epsilon_c}\right ) \biggr [ \frac{\delta(x_0-x)}{(1-y)_+} x_0 y
\nonumber\\
   &&  \ \ \ +\frac{\delta (1-y)}{(x_0-x)_+}  \frac{x(2x_0-x)}{x_0}- x_0 \delta(x_0-x)\delta (1-y) \ln\frac{q^2_\perp}{Q^2} \biggr ].
\nonumber\\
W^{(1)}_T\biggr\vert_{6c} &=& -\frac{g_s\alpha_s^2}{(4\pi)^2} \frac{N_c^2-1}{N_c} \frac{\sqrt{2x_0}}{(q^2_\perp)^2}
         \left (-\frac{2}{\epsilon_c}\right ) x_0 y (1-y)\delta(x_0-x),
\nonumber\\
W^{(1)}_T\biggr\vert_{6d} &=& \frac{g_s\alpha_s^2}{(4\pi)^2} \frac{N_c^2-1}{N_c} \frac{\sqrt{2x_0}}{(q^2_\perp)^2}
         \left (-\frac{2}{\epsilon_c}\right ) x_0 y \delta(x_0-x),
\label{Fig6}
\end{eqnarray}
the contributions from Fig.6e and Fig.6f are non-leading in the limit of $q_\perp/Q \ll 1$.
In Fig. 6a the gluon with $k_1$ corresponds to the gluon with $k_1$ in Fig.4a.
We have observed that the integration over $k_1^-$ here in Fig.6a is different that in Fig.4a.
In the lower-half complex $k_1^-$-plan there is only one pole from the quark propagator in the right part of Fig.6a,
and there are three poles from the three gluon propagators in the upper-half complex $k^-_1$-plan. One may take
a contour in the lower-half complex $k^-_1$-plan to perform the $k_1^-$-integration whose result
is only from the pole of the quark propagator. By taken a contour in the upper-half
complex $k_1^-$-plan, this integration result can also be written a sum of contributions from the three poles
of the gluon propagators. In the limit $q_\perp/Q \ll 1$, the terms in $W_T^{(1)}$ proportional to $\delta(x_0-x)$
only come from the contribution of the pole in the gluon propagator with $k_1$, i.e., these terms can equivalently
be calculated with the $k_1^-$-integration by taking Eq.(24) as for Fig.4a. With this fact
the gluon exchanged between the two gluon lines is a
Glauber gluon.  This indicates that these terms may be factorized with $T_F(x,x)$ according to the experience
from Fig.4.  The above discussed also applies for the remaining
diagrams in Fig.6. In Eq.(\ref{Fig6}) the terms proportional to $\delta(1-y)$ except the terms containing
log's can be identified as hard-pole contributions at one-loop.

\par
The sum of the diagrams in Fig.5 and Fig.6 is then
\begin{eqnarray}
W^{(1)}_T\biggr\vert_{Fig.5+Fig.6} &=& -\frac{g_s\alpha_s^2}{(4\pi)^2}  \frac{x_0\sqrt{2x_0}}{(q^2_\perp)^2}
         \left (-\frac{2}{\epsilon_c}\right )N_c (N_c^2-1) \biggr\{ \delta(x_0-x) \left [
     \frac{1+y^2}{(1-y)_+} \left ( 1+\frac{y-1}{N_c^2} \right )
\right.
\nonumber\\
   && \left. \ \  -2 \delta(1-y) \ln\frac{q^2_\perp}{Q^2} \right ] +  \frac{x\delta (1-y)}{x_0(x_0-x)_+} \biggr\}.
\end{eqnarray}
With $T_F(x,x)$ in Eq.(\ref{TFxx}) it is clearly that the terms in $ [\cdots ]$ reproduce the term ${\mathcal A}_{sq}$
in Eq.(\ref{FAC}).
The last term in the above is not a soft-gluon-pole contribution and will become relevant if one
studies the perturbative coefficient functions at the next-to-leading order.
\par
Before ending this section an interesting observation can be made. We observe
that SSA calculated here at one-loop  is generated through exchange of Glauber gluon at one-loop and
it is divergent. This is in contrast to factorizations only with twist-2 operators for Drell-Yan processes.
These factorizations are for those differential cross-sections which do not contain $T$-odd effects.
In proving these factorizations, the existence of Glauber gluons brings up the most difficult obstacle\cite{G0,G1,G2}.
But it is able to show that the divergences caused by Glauber gluons are canceled
in differential cross-sectons\cite{G0,G1,G2}. For the factorization studied here, such divergences
are not canceled and they need to be factorized into the twist-3 matrix element with $x_1 =x_2$.
This will have some implications for the study of factorizations in the framework of soft collinear effective theories
of QCD\cite{LiuMa}.

\par\vskip20pt

\par\vskip20pt
\noindent
{\bf 5. Additional Contributions}
\par\vskip10pt
In the previous sections we have used the multi-parton state in Eq.(\ref{pas}) to replace
the polarized hadron and determined the factorization form of $W_T^{(1)}$.
After the replacement, we have
studied SSA essentially in the partonic forward scattering process
$\bar q +(q +G) \to \gamma^* + X \to \bar q +q$ or the reversed scattering, where the helicity difference between the initial $qg$-state
and the final $q$-state is $\pm 1$. For a real hadron scattering, it is possible that instead of the above
scattering one has the forward scattering $\bar q + (q+\bar q) \to \gamma^* +X \to \bar q + g$
or the reversed, where the final gluon and initial $q\bar q$ come from the
polarized hadron.
If the total helicity of the $q\bar q$ state is zero, this forward scattering will also delivery
an additional contribution  to SSA besides those given in Eq.(\ref{FAC}), because the helicity  difference in the scattering
is also $\pm 1$.
This has been
realized in the study of the evolution of the twist-3 matrix element $T_F(x,x)$\cite{BMP}.
\par
The additional contribution can be factorized with the twist-3 matrix elements. The factorization can be
studied with our multi-parton state in Eq.(\ref{pas}) by adding a $qq\bar q$ state as a component:
\begin{eqnarray}
 \vert n [\lambda ] \rangle  &=&  \vert q(p,\lambda_q) [\lambda ] \rangle + c_1
                   \vert q(p_1,\lambda_q) g(k,\lambda_g ) [\lambda ] \rangle
                   + c_2 \vert q(k_1,\lambda_1)q(k_2,\lambda_2)\bar q(k_3,\lambda_3)[\lambda ]\rangle,
\nonumber\\
   k_i^\mu &=& z_i p, \ {\rm for}\ i=1,2,3.
\label{pas1}
\end{eqnarray}
Replacing the transversely polarized hadron $h_A$, one obtains SSA from the interference
of the state $qq\bar q$-state with the $qg$-state. The interference with the single quark state
will not contribute to SSA because of the helicity conservation. By taking one quark from
the $qq\bar q$-state and $qg$-state as a spectator quark, one can have the mentioned
forward scattering $\bar q + (q+\bar q) \to \gamma^* +X \to \bar q + g$ or the reversed scattering.
Therefore, we need effectively to replace, e.g., in the twist-3 matrix elements the state $\vert h_A\rangle$
with a $q\bar q$ state and the state $\langle h_A \vert$ with a gluon, or in a reversal way.
The total helicity state of the $q\bar q$ state should be zero. The $q\bar q$ state is in color octet
in correspondence with the gluon. Taking the quark with $k_2$ as the spectator, one can simply
work out those twist-3 matrix elements at tree-level:
\begin{eqnarray}
T_{F,c_2}(x_1,x_2) &=& -c_2 {\mathcal N} \pi g_s (N_c^2-1) x_0 (x_2-x_1 )
 \sqrt{2 z_1 z_3}
\nonumber\\
    && \cdot \left [   \delta(x_1+z_3) \delta (x_2-z_1)
         -\delta(x_2+z_3) \delta (x_1-z_1) \right ] ,
\nonumber\\
T_{\Delta,F,c_2}(x_1,x_2)  &=& 0.
\end{eqnarray}
The factor ${\mathcal N}$ is a normalization factor of the spectator quark state with other trivial factor.
The same factor will also appears later in $W_T^{(1)}$ and $T_F(x,x)$. All relevant quantities calculated
with the state Eq.(\ref{pas1}) are the sum of contributions studied
in previous sections and those studied here. Because of this our study of these contributions
can done separately.
It should be noted that there are two possibilities to have one quark as a spectator for the
interference with the state in Eq.(\ref{pas1}).
For simplicity we only present the results
of the above possibility in this section,
because this will not affect the factorization, i.e., the perturbative coefficient
functions. We denote quantities calculated in this way with an index $c_2$.
We have performed detailed calculations
for these two possibilities and obtained the same results for the factorization
form of $W_T^{(1)}$ and for the evolution of $T_F(x,x)$ which will be discussed later.

\par
\begin{figure}[hbt]
\begin{center}
\includegraphics[width=6cm]{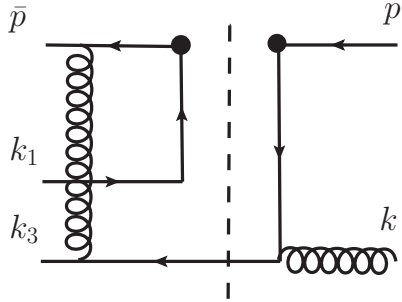}
\end{center}
\caption{The tree-level diagram contributing to SSA.}
\end{figure}
\par

Similarly, when we do the replacement of $h_A$ for the hadronic tensor as
done for the twist-3 matrix elements in the above, one can obtain nonzero SSA. At tree-level,
in the limit $q_\perp\ll Q$, only one diagram gives nonzero contribution. Calculating the contribution
in the similar way as shown in \cite{MS3} and previous sections, we obtain the structure function:
\begin{eqnarray}
W_{T,c_2}^{(1)} = c_2 {\mathcal N}  \frac{g_s\alpha_s (N_c^2-1)}{4\pi N_c^2 (q_\perp^2)^2 }
  x_0 \sqrt{2 z_1 z_3} \delta(1-y)\delta(x-z_1) \frac{ x -z_3} {x+z_3}.
\end{eqnarray}
Comparing with the above $T_F$ we can obtain a factorized form as:
\begin{eqnarray}
W^{(1)}_{T,c_2} (x,y,q_\perp)
       =\frac{\alpha_s}{(2\pi q_\perp^2)^2 N_c^2 } \int_x^1 \frac{dy_1}{y_1} \int^1_y \frac{dy_2}{y_2}
         \bar q(y_2) T_{F}(x,x-y_1)
            \delta (1-\xi_2)  (1 - 2\xi_1),
\label{Wc2}
\end{eqnarray}
it should be noted that the contribution of $T_F(x_1,x_2)$ calculated in Sect. 3 will not be involved
here because this contribution is zero for $x_2<0$. It is clear that this part should be added
to the factorized from in Eq.(\ref{FAC}), i.e., the term ${\mathcal A_h}$ should be modified as:
\begin{equation}
{\mathcal A}_h (x,y_1) =   \delta (1-\xi_2) \left [ T_F(x,y_1) \frac{1+\xi_1}{(1-\xi_1)_+}
    + T_{\Delta, F}(x,y_1)  + \frac {1-2\xi_1 }{N_c^2} T_{F}(x,x-y_1)   \right ] .
\end{equation}
In the above we note that one argument of $T_F(x_1,x_2)$ with $x_2=x-y_1$ is negative, representing
the fact that there is an antiquark from the polarized hadron.
\par
In \cite{KQ,BMP} the evolution of the twist-3 matrix elements have been studied. The results are different.
The evolution of the non-singlet part of $T_F(x,x)$ with $x>0$ is given with $z =x/\xi$ in \cite{BMP} as:
\begin{eqnarray}
\frac{\partial T_F(x,x,\mu)}{\partial \ln \mu} &=& \frac{\alpha_s}{\pi} \biggr \{
   \int_x^1 \frac{d \xi}{\xi} \left [ P_{qq}(z) T_F(\xi,\xi) + \frac{N_c}{2} \left (
   \frac{(1+z) T_F(x,\xi) -(1+z^2) T_F(\xi,\xi)}{1-z} + T_{\Delta,F}(x,\xi) \right ) \right ]
\nonumber\\
   && - N_c T(x,x) + \frac{1}{2N_c}\int_x^1 \frac{d \xi}{\xi} \left [ (1-2z) T_{F}(x,x-\xi)
      - T_{\Delta,F} (x,x-\xi ) \right ] \biggr \}.
\label{dmu}
\end{eqnarray}
The discrepancies is the following: From \cite{KQ} the evolution has only the terms in first line.
The last term in the first line has a different sign than that of \cite{BMP}.
With our multi-parton state one can calculate $T_F(x_1,x_2)$ to check the evolution.
For this we calculate $T_F(x,x)$ with the contribution from the interference with the $qq\bar q$ state
in the same way showing in the above.
\par
\par
\begin{figure}[hbt]
\begin{center}
\includegraphics[width=4cm]{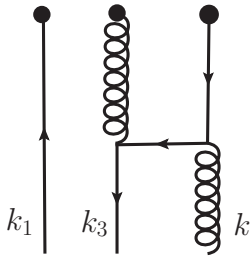}
\end{center}
\caption{The one-loop diagram to $T_{F,c_2}(x,x)$. }
\end{figure}
\par
For $x>0$ there is only one diagram at one-loop showing in Fig.8. We obtain the following:
\begin{eqnarray}
T_{F,c_2} (x,x,\mu)  &=& c_2{\mathcal N}
\frac{g_s \alpha_s (N_c^2-1)}{4N_c} x_0 \sqrt{2y_1 y_3}\delta (x-y_1)
  \frac{x-y_3}{x+y_3} \left [ -\left (\frac{2}{\epsilon_c} -\gamma +\ln 4\pi \right )
    +\ln\frac{\mu^2}{\mu_c^2}\right ].
\end{eqnarray}
Adding the results from Sect.3. we have:
\begin{eqnarray}
\frac{\partial T_F(x,x,\mu)}{\partial \ln \mu}  &=& \frac{\alpha_s }{2 \pi}
   \int_x^1 \frac{d \xi}{\xi} \left \{ N_c \left [ \frac{1+z}{1-z} T_F(x,\xi)
   + T_{\Delta,F}(x,\xi) \right ] + \frac{1}{N_c} (1-2z) T_{F}(x,x-\xi) \right \}
\nonumber\\
   && + \cdots.
\label{dmu1}
\end{eqnarray}
Comparing the above result with that in Eq.({\ref{dmu}), we find an agreement except the terms with
$T_F(x,x)$ and the last term in Eq.({\ref{dmu}). These terms are denoted with $\cdots$ in Eq.(\ref{dmu1}).
They can not be obtained at one-loop with our results here.
To verify them one has to study them
at two-loop level.
With our one-loop results at least a part of the discrepancy is solved.
\par
From the interference with the $qq\bar q$ state one also expects that
there are soft-gluon-pole contributions in Eq.(\ref{Wc2}). These contributions are generated
in a similar way as those studied in Sect. 4.  $W_{T,c_2}^{(1)}$
receives contributions at one-loop level from similar diagrams in Sect.4., where one replaces
the incoming gluon line going through the cut with an antiquark line and the outgoing quark
with an outgoing gluon line. Comparing the topology of these diagrams with
that of Fig.8 for $T_{F,c_2}(x,x)$, one can expect that the contributions to $W_{T,c_2}^{(1)}$
from those diagrams can be factorized as the soft-gluon-pole contributions in Eq.(\ref{FAC}).
We have checked these contributions. The results confirm the above expectation.

\par
\vskip20pt
\noindent
{\bf 6. Summary and Outlook}
\par\vskip10pt
Because of the helicity-conservation of QCD, SSA of a single quark state can not be generated
in the involving forward scattering.
We have used multi-parton states in order to have a nonzero SSA in the relevant partonic
processes. Using the multi-parton states one can calculate SSA of Drell-Yan processes
and relevant twist-3 matrix elements.
With these partonic results we can examine and derive the collinear factorization.
The collinear factorization for SSA has been derived in the literature based on a diagram expansion
of the relevant hadronic tensor. By using partonic results at tree-level, only the hard-pole
contributions can be obtained. The soft-gluon-pole contributions can not be obtained.
The reason for this has been discussed.
\par
In this work we have performed the study with multi-parton state at one-loop level in order
to examine and identify these soft-gluon-pole contributions in Drell-Yan processes.
If we assume that the factorization
is correct and there are only hard-pole contributions at leading order of perturbative coefficient
functions,we find that a class of one-loop
contributions to SSA, which are collinearly divergent, can not be factorized
at one-loop order.
To correctly factorize collinear divergences
appearing in these contributions, one has to add extra terms in the factorization
derived with tree-level partonic results. Interestingly these extra terms are just those soft-gluon-pole
contributions. Therefore, with our multi-parton states we can re-derive the soft-gluon-pole contributions
which have been derived with the diagram expansion in\cite{JQVY1}.
\par
It is interesting to note that the hard-pole- and soft-gluon-pole contributions in SSA, i.e., in the
structure function $W_T^{(1)}$ are at different order of $\alpha_s$. But, the perturbative coefficient
functions are at the same order.
The perturbative coefficient functions associated with the soft-gluon-pole contributions,
although derived from SSA at one-loop, are at the same order of those associated with the hard-pole contributions,
which are derived from SSA at tree-level. This is in contrast with factorizations for differential cross-sections,
where only leading twist-2 operators are involved. In these twits-2 factorizations,
perturbative coefficient functions at leading order are completely determined by differential cross-sections
at tree-level and tree-level matrix elements of involved twist-2 operators. Therefore, our study here also shows
an unusual feature of factorizations involving twist-3 operators.
\par
By taking multi-parton state we also find a new contribution to $W_T^{(1)}$. The new contribution
comes from the parton process where a $q\bar q$ pair with the total helicity $\lambda =0$
is transmitted into a gluon. This new contribution can be factorized with the twist-3
matrix element.
One important twist-3 matrix element is calculated at one-loop in this work.
From the result one can derive the evolution of the matrix element.
In this work we can solve
a part of discrepancy between evolutions derived in \cite{KQ,BMP}. To solve
the remaining parts one has to calculate the twist-3 matrix elements with the multi-parton state
at two-loop.
\par
In this work we have restrict ourself to the certain relevant partonic processes with
one antiquark from the unpolarized hadron.
Having succeeded to reproduce the soft-pole-gluon pole contributions with multi-parton states,
one can start to analyze with the method presented in this work other relevant partonic processes.
E.g., the processes involving one gluon
from the unpolarized hadron. In such processes, it is possible to have soft-quark-pole contributions
represented by $T_F(0,x)$. One can also use our method to analyze another type of SSA appearing in the case
when the momentum of each lepton is measured. In this case the factorization of SSA
is so far still different derived from different works\cite{SSA0QT}. In principle, our method
is not restricted to Drell-Yan processes. It can be used in any case where SSA appears.
 We believe that it has advantages to use our method
for analyzing factorizations of SSA and for calculating higher order corrections, because the involved
calculations are of standard scattering amplitudes.

\par
\vskip 5mm
\par\noindent
{\bf Acknowledgments}
\par
This work is supported by National Nature
Science Foundation of P.R. China(No. 10975169,11021092).
\par\vskip30pt

\par

\end{document}